%% file: main.tex
\documentclass[conference]{IEEEtran}
\usepackage{amsmath}

\usepackage{amssymb,amsfonts}
\usepackage{algorithmic}
\usepackage{graphicx}
\usepackage{textcomp}
\usepackage{xcolor}
\usepackage{multirow}
\usepackage{array}
\usepackage{booktabs}
\usepackage{caption}
\usepackage{subcaption}
\usepackage{threeparttable}
\usepackage{marvosym}
\input{self_defined_command}

\begin{document}
\title{DPUV4E: High-Throughput DPU Architecture Design for CNN on Versal ACAP}

\author{\IEEEauthorblockN{Guoyu Li\textsuperscript{*}, Pengbo Zheng\textsuperscript{\Letter}, Jian Weng and Enshan Yang}
\IEEEauthorblockA{Advanced Micro Devices, Inc. \\ \{guoyuli2, pengbo.zheng, jian.weng, enshan.yang\}@amd.com}
}

\maketitle

\newcommand\blfootnote[1]{%
\begingroup
\renewcommand\thefootnote{}\footnote{#1}%
\addtocounter{footnote}{-1}%
\endgroup
}
\blfootnote{
\hspace{-0.04in}\text{$*$} \text{Contribution during internship at AMD} \\
\text{
\Letter} \text{Corresponding author}
}

\input{0.abstract}

\input{1.introduction}
\input{2.relatedwork}

\input{3.architecture}

\input{4.aie-design}

\input{6.pl-design}

\input{7.evaulation}

\input{8.conclusion}
\input{main.bbl}

\end{document}

%% file: self_defined_command.tex
\usepackage{xcolor}

\newcommand{\myparagraph}[1]{\textbf{\textbf{#1 }}}

%% file: 0.abstract.tex
\begin{abstract}
Convolutional Neural Networks (CNNs) remain prevalent in computer vision applications, and FPGAs, known for their flexibility and energy efficiency, have become essential components in heterogeneous acceleration systems. However, traditional FPGAs face challenges in balancing performance and versatility due to limited on-chip resources.
AMD’s Versal ACAP architecture, tailored for AI applications, incorporates AI Engines (AIEs) to deliver high computational power. Nevertheless, the platform suffers from insufficient memory bandwidth, hindering the full utilization of the AIEs’ theoretical performance.
In this paper, we present DPUV4E for the Versal architecture, providing configurations ranging from 2PE ($32.6$ TOPS) to 8PE ($131.0$ TOPS). We design two computation units, Conv PE and DWC PE, to support different computational patterns.
Each computation unit’s data flow efficiently utilizes the data reuse opportunities to mitigate bandwidth bottlenecks. Additionally, we extend the functionality of each PE to utilize AIEs for non-convolutional operations, reducing resource overhead.
Experiments on over 50 models show that compared to previous designs, our design provides $8.6\times$ the TOPS/W of traditional FPGA-based DPU designs, while reducing DSP usage by $95.8\%$, LUT usage by $44.7\%$, and latency to $68.5\%$ under single-batch conditions. For end-to-end inference, our design improving throughput by up to $2.2\times$ for depth-wise convolution models and up to $1.3\times$ for standard models. 
\end{abstract}

%% file: 1.introduction.tex
\section{Introduction}
At present, deep learning (DL) has profoundly integrated into our daily lives. Despite the emergence of new transformer-based neural networks, Convolutional Neural Networks (CNN) remain extensively employed owing to their proficiency in extracting local information from images in relatively smaller datasets.
GPUs' efficient parallel processing is used to improve CNN inference, but their general-purpose design reduces energy efficiency. To improve accelerators' energy efficiency and throughput, custom CNN architectures have been proposed.

Nonetheless, the expense of standalone ASICs is substantial, whereas FPGAs, due to their reconfigurability, superior energy efficiency, and lower costs, serve as the preferred implementation platform for numerous accelerator prototypes.
As HLS gains popularity, software writers can efficiently transition particular activities to the CPU+FPGA heterogeneous system, resulting in significant performance enhancements.

To address the growing demand for heterogeneous computing, Xilinx has developed Versal Adaptive Compute Acceleration Platform (Versal ACAP). Versal ACAP integrates traditional programmable logic (PL), software processor (PS), and a high-performance AI Engine. 
The Xilinx team introduced the configurable DPU design XVDPU\cite{xvdpu} for the ACAP platform, which can flexibly generate DPU overlays on demand. In VCK190, it supports the use of 96 to 320 AIE Cores to achieve peak performance of 32.76 to 109.2TOPS, and can run the ResNet50 network at 1653 to 4050FPS. Compared to DPU designs solely based on FPGA \cite{DPUCZDX8G,DBLP:journals/tnn/LiuFFNSL22,h2pipe}, this boosts throughput by 6.2 $\sim$ 24 $\times$. 
However, this design does not fully utilize the datapath capabilities provided by AIE and exhibits low efficiency in managing Depthwise Convolution (DWC) operations, which are commonly encountered in neural networks. Furthermore, the XVDPU employs Programmable Logic (PL) to execute element-wise operations and activations, leading to substantial consumption of valuable DSP resources. This limitation hinders the integration of additional custom logic by users.

To further enhance the high-throughput requirements of the DPU design based on the AI Engine, we have designed DPUV4E. Our main contributions are:

\begin{itemize}
  \item We designed and implemented a high-throughput DPUV4E on the Xilinx ACAP, efficiently conducting AI inference on the FPGA end using the AI Engine. It facilitates end-to-end inference of hundreds of CNN models and provides $8.6\times/1.4\times$ TOPS/W improvement compares to FPGA-based and AIE-based DPU designs.
  \item We designed an innovative convolution datapath that makes full use of the AIE cascade channel for data accumulation and leverages AIE resources for element-wise and activation operations. Under the same batch configurations, our DSP/LUT/FF consumption amounts to only $4.2\%$, $55.3\%$, and $56.5\%$, respectively, compared to previous work.
  \item We have devised a specialized DWC engine that alleviates the problem of low utilization caused by DWC operations during CNN inference while also being compatible with standard convolution operations. It delivers a performance improvement ranging from $1.3\times$ to $2.1\times$ for DWC models without affecting the performance of traditional models.
  \item An optional Low-Channel Conv Engine in PL to accelerate the initial stage of CV inference, providing a performance boost of $1.14\times$ on ResNet50 compared to the normal design.
  \item The methodology from parallelism modeling to design space exploration on the Versal platform to efficiently leverage the full memory bandwidth of the AIE. Through fine-grained engineering optimization, this approach facilitates the operation of the entire design at high frequencies with enhanced efficiency.
\end{itemize}

%% file: 2.relatedwork.tex
\begin{figure}[t]
    \centering
    \includegraphics[width=0.8\columnwidth]{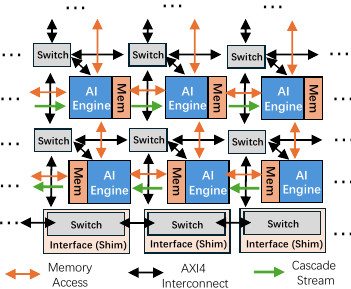}
    \caption{AI Engine Architecture}
    \label{fig:aiengine}
    \vspace{-0.15in}
\end{figure}

\section{Background}
\subsection{Versal ACAP}
\label{sec:background-aie}
\label{sec:versal_acap_architecture}
The Versal ACAP architecture incorporates:
(1) \textbf{Programmable Logic (PL)}: Traditional FPGA logic with DSP, allowing users to implement custom logic to accelerate domain-specific tasks.
(2) \textbf{Processing System (PS)}: A hard-core ARM processor capable of running an operating system and controlling the execution of the PL design through AXI.
(3) \textbf{AI Engine (AIE)}: The AMD VC1902 chip features a first-generation AI Engine, comprising an array of $8\times50$ AIE Tiles. Each AIE includes a high-performance SIMD processor adopting VLIW, with a vector unit capable of performing up to $128$ INT8 MAC operations per cycle. Additionally, each AIE has two load units and one store unit for concurrent vector access. The AI Engine offers a variety of intrinsics and is released as a C++ library, enabling users to write AI Engine high performance kernel code using C++.

The AIE, PL, and PS are interconnected via a Network on Chip (NoC), which also connects to other I/O peripherals such as the PCI-E/DDR4-DIMM Controller. The AMD VCK5000 Card with VC1902 chip could provide a PL $\leftrightarrow$ DDR4-DIMM bandwidth of $102.4$ GB/s.
\begin{figure*}[ht]
    \centering
    \includegraphics[width=0.7\textwidth]{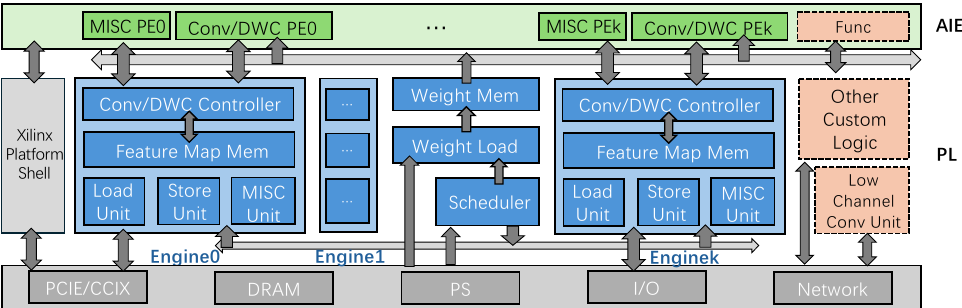}
    \caption{DPUV4E Overview}
    \label{fig:dpuv4e}
    \vspace{-0.2in}
\end{figure*}

\vspace{-0.1in}
\subsection{Data path in AIE}
Figure \ref{fig:aiengine} illustrates the basic architecture and data path of AI-Engine\cite{aiearch}.

\label{sec:background-aiedatapath}
\myparagraph{AIE$\leftrightarrow$PL} The data transmission between the AIE and PL requires clock domain crossing through $50$ AIE Interface Tiles located beneath the AIE array, only $39$ of these Interface Tiles have PL connections. Through the Interface Tiles, the PL can establish up to $8$ PL$\rightarrow$AIE streams and $6$ AIE$\rightarrow$PL streams, each with a width of $64$ bits.
These Interface Tiles can provide up to $1.3$ TB/s of PL$\rightarrow$AIE bandwidth and $1.0$ TB/s of AIE$\rightarrow$PL bandwidth in total, which is $10\times$ the off-chip bandwidth. This necessitates efficient data reuse by designers to maximize the coverage of communication latency.

\noindent\myparagraph{AIE$\leftrightarrow$AIE} The AIE array offers multiple memory access paths. Each AIE Core is equipped with a 4MB Memory Module organized as $8$ banks, each containing $256$ words with a width of $16$ bytes. Adjacent AIEs can directly access the data memory of their neighbors via AXI interfaces or transmit data directly through dedicated streaming channels. Non-adjacent AIEs can interconnect through a NoC constructed with AXI-Stream switches. 

\vspace{-0.05in}
\subsection{FPGA-Based AI Accelerator}
The FPGA is an excellent choice for heterogeneous AI inference acceleration due to its customizability and high energy efficiency. To fully leverage the FPGA’s customizability, DNNBuilder\cite{dnnbuilder} and Deepburning\cite{deepburning} assign dedicated accelerators for each layer and organize them into a fully pipelined structure. H2PIPE\cite{h2pipe} represents the current state-of-the-art accelerator based solely on FPGA logic, utilizing High Bandwidth Memory (HBM) to enhance the inference of large-scale CNN models.
DNNExplorer\cite{dnnexplorer} and Deepburning-SEG\cite{deepburningseg} employ design space exploration (DSE) to identify a limited number of accelerator units that follow different computation patterns. To support a diverse array of models, both the AMD DPU \cite{DPUCZDX8G} and NVDLA \cite{nvdla} are designed with efficient architectures specifically optimized for general operators, thereby accelerating model inference.

The Versal platform introduces high-performance AIEs for vector operation acceleration on the traditional FPGA platform, creating new opportunities for neural network inference on FPGAs.
Recent works\cite{maxeva}\cite{DBLP:conf/dac/ZhuangYZ23}\cite{DBLP:conf/fccm/TakaGGMA24} employ DSE to investigate methods for high-performance General Matrix Multiply (GEMM) using a single scale on the AIE.
CHARM\cite{charm} further explores GEMM computation strategies across multiple computational scales and proposes a code generation framework. The XVDPU\cite{xvdpu} is the only end-to-end AI inference accelerator within the Versal ACAP architecture, employing a fixed-size compute array to implement various model inferences.

%% file: 3.architecture.tex
\vspace{-0.05in}
\section{DPUV4E Architecture}
\vspace{-0.05in}
\subsection{Overview}
The detailed hardware architecture of the DPUV4E is shown in the Figure \ref{fig:dpuv4e}. 
To support different devices, DPUV4E can be configured to instantiate 2/4/6/8 Computing Engines and various Buffer Sizes according to FPGA resource constraints.

Each Computing Engine is divided into two parts: data buffers and control logics are implemented on PL side using CLBs, and PE is located on the AIE. 
Depending on the workload, DPUV4E provides two types of Computing Engines: Conv PE(only support Conv) or DWC PE (support DWC+Conv). We also offer a MISC Core, optionally implemented in AIE/PL, to perform non-convolution operations such as element-wise addition and pooling. Implementing MISC in AIE can significantly conserve PL DSP usage, thereby allowing for the implementation of more custom optimization logic.
All computing engines share one scheduler unit, implemented in PL. To maximize the computational capabilities of AIE, the model must be quantized to INT8 and executed on the DPUV4E.

The DPUV4E uses on-chip buffers to buffer weights, biases, and intermediate values to increase on-chip data reuse. Each computing engine has independent feature map buffer and shares same weight buffer. DRAM is used as system memory to store instructions, input images and inference results. After initialization, the DPU controller fetches instructions from off-chip DDR4 to control the operations of the computing engine.

%% file: 4.aie-design.tex
\vspace{-0.05in}
\section{AI-Engine Level Design}
\vspace{-0.05in}
In this section, we will introduce the specification analysis and detailed design of the DPUV4E in AIE part.
We use IH/IW/OH/OW/IC/OC to represent the height/width/channel dimensions of the input/output feature maps, respectively.
\vspace{-0.1in}
\subsection{Parallelism Modeling}
\label{sec:aie-analysis}
Although AIE-PL provides much higher bandwidth than off-chip sources, it remains bandwidth-limited in convolution scenarios. 
The AIE MAC employs a shared operand architectural design, optimizing efficiency for $1\times16\times8$ GEMM operations\cite{aie-intrinsic}. That requires a single AIE to have $1024$ bits of data input and $64$ bits of output. However, as mentioned in Section \ref{sec:background-aiedatapath}, the $400$ AIEs are served by only $39$ AIE Interface Tiles, each of which can provide only a $192$-bit input stream and a $128$-bit output stream, resulting in a severe imbalance. 

To address the bandwidth bottleneck, DPUV4E conducts design space exploration based on allocation of limited I/O bandwidth to find the necessary levels of data reuse under different bandwidth allocations for feature maps and weights, thereby defining the design specifications. 
In $1\times16\times8$ INT8 MAC, the three dimensions correspond to one pixel of the convolution, the input channels (IC), and the output channels (OC).
This computation scale results in a pixel size = 1, allowing the influence of the input feature map size to be ignored during single iteration modeling, making the MAC kernel design more generalizable. 

In most cases, convolution operations encounter memory bottlenecks, necessitating the prioritization of determining the bandwidth required for loading weights ($BW_w$) and feature maps ($BW_f$). To optimally balance computation and storage time, it is desirable to:
\begin{equation}
    \begin{aligned}
        &\left\{\begin{array}{l}
\text{FMLoad} = \text{WTReuse} \times 16 \times 8 / BW_f
 \\
\text{WTLoad} = 16\times 8\times 8\times  \text{FMReuse}/BW_w
 \\
T_{mac}  = \text{WTReuse}*\text{FMReuse}
\end{array}\right.
\\
&s.t.
\left\{\begin{array}{l}
\text{FMLoad} \le  T_{mac}
\\
\text{WTLoad} \le T_{mac}
\end{array}\right.
    \end{aligned}
\end{equation}
For example, consider FMLoad, which denotes the cycles required to load feature maps. Its value is calculated as the product of the reuse count of each weight vector ($1\times16$) and the length of each feature vector ($8\times16$), divided by the bandwidth ($BW_f$). Meanwhile, the total computation time ($T_{mac}$) is numerically equivalent to the reuse product count.

Here, FMReuse in convolution refers to the results corresponding to different kernels, necessitating the loading of multiple weights to extend the OC dimension. WTReuse requires loading the sliding window of the input image associated with one weight vector to extend the pixel size. Therefore, to achieve such an ideal reuse rate, we aim for the entire convolution task to attain:

\begin{equation}
    \begin{aligned}
        \left\{\begin{array}{l}
\text{OC} \ge 8 \times \text{FMReuse}
 \\
\text{IH} \times \text{IW}  \ge \text{WTReuse}
\end{array}\right.
    \end{aligned}
\end{equation}

\input{tables/reuse-required}
We can determine the minimum parallelism scheme that achieves $CTC=1$ under different conditions $BW_f$ and $BW_w$, as shown in Table \ref{tab:reuse-required}. DPUV4E selects the scheme with $BW_f=32\text{-bit}$ and $BW_w=16\text{-bit}$ for subsequent Conv PE design, as it represents a favorable compromise. The configuration with OC=32 achieves high utilization across hidden layers in various models, and the design with IH$\times$IW=64 effectively utilizes the size of local memory in the ACC/NL Core to temporarily store partial sums (detailed in Section \ref{sec:convpe-accnl}).

\vspace{-0.05in}
\subsection{Convolution PE} 
\label{sec:conv-pe}
\begin{figure}[hbt]
    \centering
    \includegraphics[width=0.9\columnwidth]{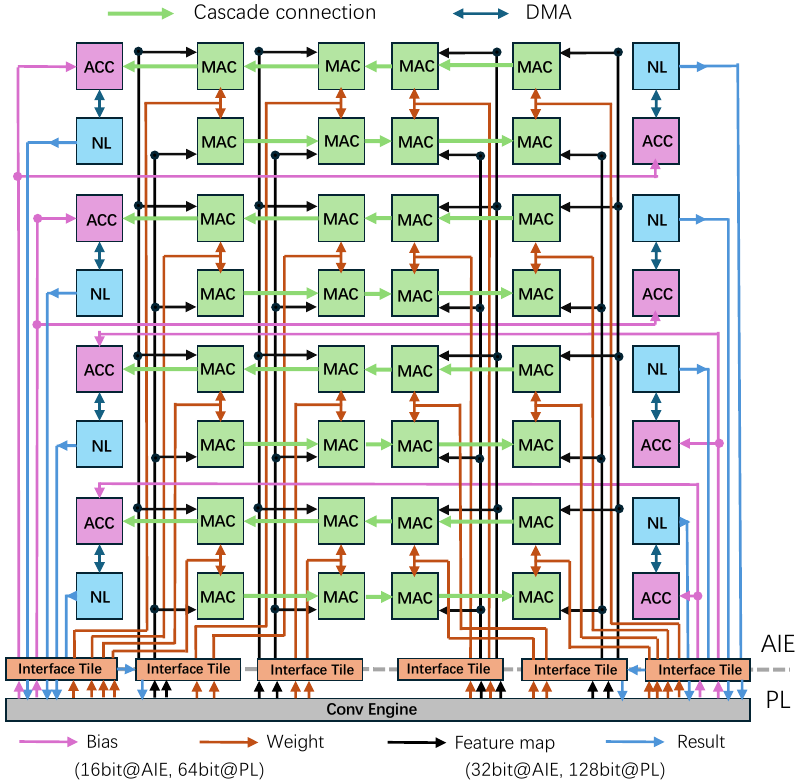}
    \caption{PE for Convolution Calculation}
    \label{fig:conv-pe}
    \vspace{-0.2in}
\end{figure}
The convolution PE is used to handle computation-intensive operations, including traditional convolution and linear operations. Figure \ref{fig:conv-pe} illustrates the architecture of the AIE array of convolution PE. 
48 AIE cores organized in an 8-row by 6-column grid. The central 4 columns are dedicated to multiply accumulate (MAC) operations, while the outer columns on both sides handle accumulation (ACC) and nonlinear (NL) operations.

\subsubsection{MAC Core}
\begin{figure}[hbt]
    \centering
    \includegraphics[width=\columnwidth]{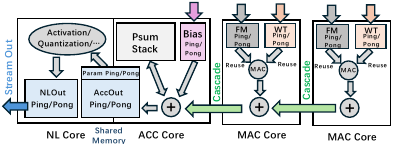}
    \caption{MAC-ACC-NL Chain Kernel Design}
    \label{fig:mac-kernel}
    \label{fig:acc-kernel}
\end{figure}

Each MAC Core is responsible for computing a $1(IH/IW)\times16(IC)\times8(OC)$ matrix multiplication. 
Each cycle produces 8 $48-$bit results, which are sent to the next MAC Core via a $384$-bit cascade stream interface. The next MAC Core performs element-wise accumulation in the cascade pipeline and starts the next cascade transfer, achieving in-flight accumulation during data transfer.

As stated in Section \ref{sec:aie-analysis}, adopting this computational scale necessitates meeting the constraints $OC \ge 32$ and $IH \times IW \ge 64$. These constraints can be realized across multiple dimensions. In DPUV4E, we choose to reuse the weight $4$ times at the Core-level, meaning each MAC Core computes $4(IH)\times16(IC)\times32(OC)$ using $16$ cycles before proceeding with cascade transmission. Please note that IH is not arbitrarily chosen; multiple computation results of the MAC Core need to be accumulated in the ACC Core before being written back to PL. The selection of IH and IW is constrained by the storage capacity that the ACC Core provides for partial sums, as detailed in Section \ref{sec:convpe-accnl}.

\subsubsection{ACC\&NL Core}
\label{sec:convpe-accnl}
The ACCumulation (ACC) and Non-Linear (NL) Core are connected at the end of the MAC chain. ACC core store the partial sums, performing bias and kernel-level accumulation. NL core will then perform non-linear activation, element-wise operations and quantization etc. to generate output feature maps. Unlike previous designs \cite{xvdpu} where such operations are implemented using DSPs on the PL side, DPUV4E implements accumulation and activation using separate AIE Cores to save resources.

Figure \ref{fig:acc-kernel} shows the data flow design of the ACC and NL. 
Before the MAC Chain completes a full iteration, the ACC Core must store the partial sum vectors output from the MAC Chain each cycle and process element-wise accumulate when necessary. The size of the partial sum buffer that ACC Core provides is the primary bottleneck during this process. Memory allocation also needs to consider the banking division of data to prevent potential bank conflicts.

According to \cite{aiearch}, each AIE Core's local memory consists of 8 banks, each sized at $256\times128$-bit, totaling $32$KB. The AIE can access adjacent memory modules, providing us with 64KB of storage space across 16 banks for an ACC-NL pair. The primary storage overhead is in PsumStack (storing intermediate data for partial sum accumulation), AccOutBuffer (storing the feature map after accumulation), BiasBuffer (storing the bias), and NLOutBuffer (storing the activation data). To match the read data volume in a single accumulation process, we also require the buffers used in the same cycle to have the same read/write bandwidth (mapped to the number of banks). Therefore, we have:
\begin{equation}
\hspace{-0.4cm}
\begin{aligned}
&\left\{\begin{array}{l}
\text{PsumStack} = \text{AccOut} = \text{IH}\times\text{IW}\times32\text{(OC)}\times4B
 \\
\text{BiasBuf} = 32\text{(OC)} \times 4B
\\
\text{NLOutBuf} = \text{IH} \times \text{IW} \times 32\text{(OC)} \times 1B
\\
\text{Bank}_{\text{PsumStack}} = \text{Bank}_{\text{AccOut}} = \text{Bank}_{\text{Bias}} = \text{Bank}_{\text{NLOut}}
\end{array}\right.
\\
&s.t. \\
&\left\{\begin{array}{l}
\hspace{-0.3cm}
\quad \text{PsumStack} + 2 \times \text{(Other Buffer Size)} \le 64KB \\
\hspace{-0.3cm}
\quad \text{Bank}_{\text{PsumStack}} +  2 \times \text{(Other Buffer Banks)} \le 16
\end{array}\right.
\end{aligned}
\end{equation}

In the equation, the factor of $2$ represents the use of PingPong buffer, and 
4B and 1B denote the intermediate accumulation and output bitwidth, respectively.
We can easily conclude that each buffer requires two banks to balance computation and data aggregation. Among all Banks, the one that occupies the most space is the PSumStack. As discussed in Section \ref{sec:aie-analysis}, we set IH=$4$ for each MAC Core, we can adjust IW to accommodate the storage size. The constraints can be expressed as:

\vspace{-0.05in}
\begin{equation}
\begin{aligned}
\text{AccOutBuf} &= 4 \times \text{IW} \times 32 \times 4B \le 2 \times 8KB
\\
\therefore \text{IW}_{max} &\le 32
\end{aligned}
\end{equation}

When IH=4, to satisfy the ACC partial sum storage requirements and the overall data reuse constraint (IH$\times$IW $\le$ 64) simultaneously, we set the IW=16.

\subsubsection{Graph-Level Design}
\label{sec:convpe-graphlevel}
The data reuse discussed above only leveraged AIE-level reuse. In the DPUV4E, we also utilized graph-level data reuse. By using the AIE$\leftrightarrow$AIE Cascade connection, adjacent MAC Cores process different IC tile from the same input feature map pixel and accumulate them to enhance graph-level IC parallelism. Additionally, we employ stream multicast to broadcast the input feature map and weight, thereby extending the IH and OC dimensions.

As illustrated in Figure \ref{fig:conv-pe}, each input weight/offset is reused by $2$ MAC Cores, resulting in an increase in the IH dimension per iteration from 4 to 8. Each input feature map is similarly broadcast to $4$ MAC Cores, extending the OC dimension to $32\times4=128$. By means of cascading, a sequence of $4$ MACs can concurrently compute \(16 \times 4 = 64\) ICs and aggregate them into a vector of length 16. Therefore, the iteration granularity of DPUV4E Conv PE at the graph level is $8(IH)\times64(IC)\times128(OC)$.

\subsubsection{Cascade pipeline efficiency}
\begin{figure}[bt]
    \centering
    \includegraphics[width=\columnwidth]{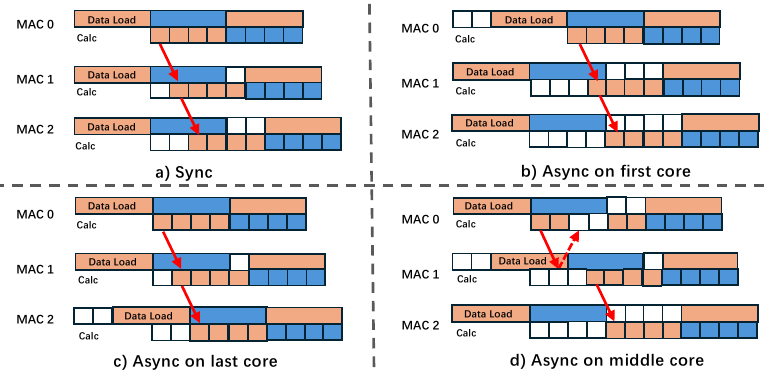}
    \caption{Automatic elimination of bubbles in the datapath}
    \label{fig:cascade-pipeline}
    \vspace{-0.25in}
\end{figure}
The MAC Chain design, which utilizes a cascade channel, necessitates the synchronization of all MAC Cores at runtime to prevent errors in IC accumulation. As each AIE’s data transfer is asynchronous, in some cases this could affect the operation of the entire chain due to data dependencies. Previous works\cite{xvdpu} only involved data transfer between two MAC Cores, thus addressing this issue was not critical.
DPUV4E exacerbates losses from pipeline bubbles by linking more MAC cores. We have devised an automated bubble elimination scheme to efficiently utilize the cascade pipeline.

Figure \ref{fig:cascade-pipeline} depicts the asynchronous I/O scenarios that may arise in the head, tail, and middle Cores of a MAC Chain. Typically, we configure each AIE Kernel Code to introduce a one-cycle delay and prohibit reading from the cascade channel during the initial phase. When pipeline bubbles appear at the head and tail of the MAC Chain, they only result in additional blocks during the initial stage.
When a pipeline bubble occurs in the middle, as illustrated in Figure \ref{fig:cascade-pipeline} d), MAC1 transmits the calculation result to MAC2, while MAC2 has not entered the data reading phase. The backpressure generated by MAC2 in the cascade channel will impede the subsequent writing of results by upstream AIEs, while downstream AIEs will remain in a waiting state owing to read obstruction from the cascade channel, until MAC2 finalizes its data I/O. After completion of the initial phase, all subsequent transmissions can guarantee 100\% usage of the MAC Chain.

\subsection{Depth-wise Convolution PE}
\begin{figure}[bt]
    \centering
    \includegraphics[width=0.9\columnwidth]{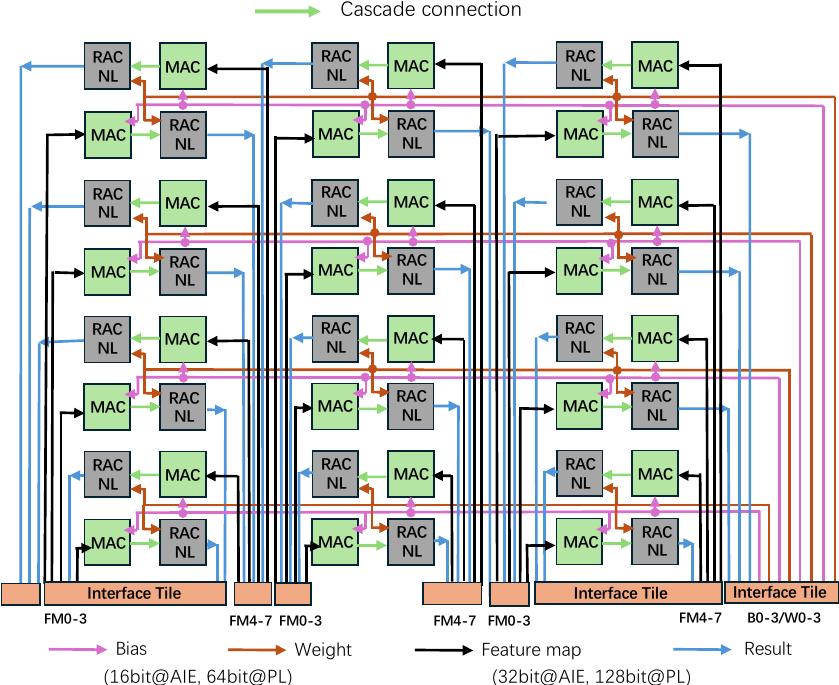}
    \caption{PE for Depth-wise Convolution}
    \label{fig:conv-dwc}
\end{figure}

Depthwise Separable Convolutions\cite{Xception} possess a reduced number of parameters and execute fewer operations compared to regular convolutions, and they are employed in numerous CNN architectures and vision transformers\cite{vit} (ViTs) to mitigate weight overhead. It comprises depthwise and pointwise convolutions. In depthwise convolution (DWC), each convolutional kernel operates exclusively on a single channel. This indicates that AIE's MAC cascade cannot leverage parallelism in the input channel dimension. 
Moreover, the sharing of input feature maps among several kernels precludes on-chip data reuse. 
Previous research \cite{dwc-opt1,dwc-opt2,dwc-opt3} has investigated the acceleration of DWC on general-purpose GPUs and FPGAs; however, no optimization strategies for AIE have been explored. 
DPUV4E provides a novel processing element optimized for DWC operations on AIE, along with the construction of specialized data flows to enhance hardware usage.

Figure \ref{fig:conv-dwc} illustrates the DWC PE's AIE array design. 
The 48 AIEs are stratified into 3 groups, each comprising 8  MAC-RACNL pairs. The MAC core executes MAC operations, while the RACNL core executing results accumulation, non-linear operations, and quantization. Please note that the DWC PE and the Conv PE have the same number of AIE Tile; however, the DWC PE requires additional PL Interface. This design allows the DWC PE \textbf{compatible} with traditional convolution operations. Nevertheless, due to the limited availability of PL Interfaces, a DPUV4E design utilizing DWC PE can support a maximum of 6PE, compared to 8PE for the Conv PE.

\subsubsection{MAC-RACNL Pair}

To efficiently leverage the AIE’s $(1\times16)\times(16\times8)$ matrix multiplication parallelism necessitates intricate rearrangement of the feature map.
This requires further hardware logic in the PL and is exceedingly time-intensive.
In the DWC MAC core, we choose to directly segment the feature map, with each core calculating the convolution outcomes for a single feature map tile and the complete kernel.

Under this strategy, only 2 sets of int8 multiplications, each with two 16-length vectors, can be calculated simultaneously due to hardware limitations. 
To facilitate data bursts, zero padding is applied to the relevant pixels of the weight data according to the convolution stride, and all data is aligned to the bank width (16B). The memory module contains two banks for both weights and feature maps, enabling the concurrent reading of four vectors, each of length 16, necessary for computation.

\begin{figure}[t]
    \centering
    \includegraphics[width=0.65\columnwidth]{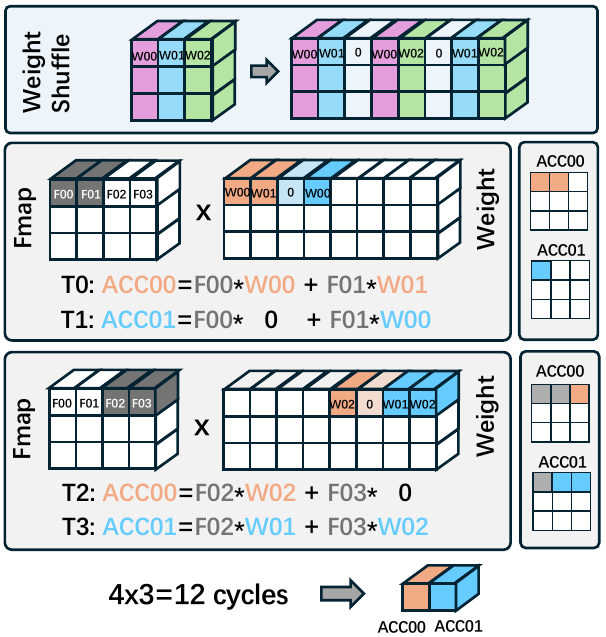}
    \caption{Depth-wise Convolution Data flow}
    \label{fig:dwc-dataflow}
    \vspace{-0.2in}
\end{figure}

Figure \ref{fig:dwc-dataflow} depicts the sequence of computations for a kernel size of $3$ and a stride of $1$. A $3\times3$ kernel necessitates two cycles to retrieve three weight vectors. The physical storage pre-inserts zero data into the weight data required for odd-numbered computations, ensuring that the two required vectors always have the same word index in the memory bank, thus streamlining the branching overhead in AIE programming. As shown in Figure \ref{fig:dwc-dataflow}, producing $1(OH)\times2(OW)\times16(C)$ accumulated result requires $12$ cycles, referred to as one atomic DWC computation.
\begin{figure}[t]
    \centering
    \includegraphics[width=0.8\columnwidth]{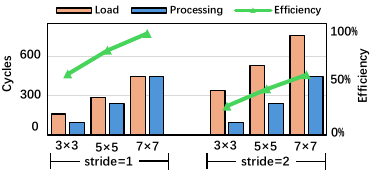}
    \caption{DWC PE data load and processing time under different convolution parameters}
    \label{fig:dwc-efficiency}
    \vspace{-0.25in}
\end{figure}
An atomic DWC computation exhibits varying input fmap dimensions and computation cycles based on the kernel size and stride. To optimize the on-chip weight data, we configure the output dimensions of each MAC-RACNL Chain iteration to $2(OH)\times8(OW)\times16(OC)$, equating to $8$ atomic DWC calculations. Figure \ref{fig:dwc-efficiency} presents the data reading and processing duration necessary for one iteration with standard convolution parameters.

As shown in Figure \ref{fig:dwc-efficiency}, when $stride=1$, FM input is the primary bottleneck, precluding the attainment of a 1:1 CTC ratio as observed in Conv PE. In DWC PE, a larger convolution kernel amplifies the computational density on the feature map, leading to an elevated CTC ratio, whereas an increased stride augments the computational sparsity on the feature map, resulting in a diminished CTC ratio. When $7\times7$ DWC is performed, the design achieves the highest efficiency.

\subsubsection{Graph-level Design}
The key focus of graph-level design is on the reuse of weights. At the graph level, as illustrated in Figure \ref{fig:conv-dwc}, every 2 rows of AIE Cores constitute a cluster, sharing a single weights port and a single bias port. The MAC cores that share weights possess distinct feature map tile inputs, executing block-wise depth-wise convolution operations. According to \cite{aiearch}, each AIE Core supports two 32-bit AXI-Stream channels for inputting data from the PL, allocated to weight and feature map input, respectively.
Compared to Conv PE, the output channels of DWC PE have increased by $3\times$ (24 channels), making it impossible to output results through the 6 AIE Memory Interfaces of the 6 columns of Cores (maximum of 18 output channels). Consequently, the architecture of DWC PE utilizes neighboring Interface Tiles for output, resulting in certain layout constraints that enable DWC PE.

%% file: tables/reuse-required.tex
\begin{table}[]
\centering
\caption{Data reuse requirements under different bandwidths}
\vspace{-0.05in}
\label{tab:reuse-required}
  \setlength{\tabcolsep}{3pt} %
  \renewcommand{\arraystretch}{1.2} %
\begin{tabular}{ccccccc}
\hline
\textbf{Inputs}                                                                         & \boldmath{$BW_f$} & \boldmath{$BW_w$} & \textbf{FMReuse} & \textbf{WTReuse} & \textbf{OC}     & \textbf{IH $\times$ IW} \\ \hline
\multirow{4}{*}{\begin{tabular}[c]{@{}c@{}}FM:\\ (1$\times$16) \\ WT:\\ (16$\times$8)\end{tabular}} & 16              & 16              & $\le$ 8            & $\le$ 64           & $\le$ 64          & $\le$ 64                  \\ \cline{2-7} 
                                                                                        & 16              & 32              & $\le$ 8            & $\le$ 32           & $\le$ 64          & $\le$ 32                  \\ \cline{2-7} 
                                                                                        & \textbf{32}     & \textbf{16}     & \textbf{$\le$ 4}   & \textbf{$\le$ 64}  & \textbf{$\le$ 32} & \textbf{$\le$ 64}         \\ \cline{2-7} 
                                                                                        & 32              & 32              & $\le$ 4            & $\le$ 32           & $\le$ 32          & $\le$ 32                  \\ \hline
\end{tabular}
\vspace{-0.2in}
\end{table}

%% file: 6.pl-design.tex
\vspace{-0.05in}
\section{PL Design}
The AIE of DPUV4E offers an independent architecture essential for computation, while traditional data preprocessing, instruction decoding, convolution control, and DDR memory access are handled on the PL side. 
\vspace{-0.05in}
\subsection{Convolution Engine}
The Convolution Engine decodes model files compiled by the host and executes memory I/Os. The architecture of the Convolution Engine is depicted in Figure \ref{fig:dpuv4e}. The Convolution Engine comprises an FM Buffer, a Conv Controller, a MISC Unit, and a Load/Store Unit.
\begin{figure}[bt]
    \centering
    \includegraphics[width=0.8\columnwidth]{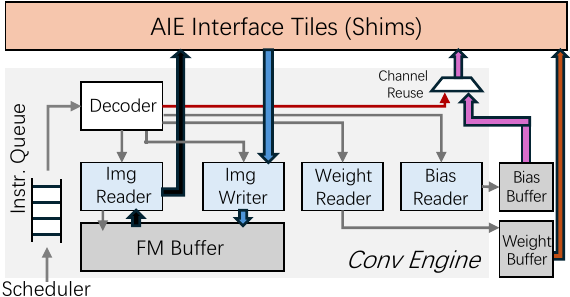}
    \caption{Convolution controller}
    \label{fig:convcontroller-arch}
    \vspace{-0.25in}
\end{figure}
The convolution controller architecture, illustrated in Figure \ref{fig:convcontroller-arch}, is designed to efficiently manage the convolution process. The controller acquires instructions from the instruction queue within the Scheduler module and, after decoding, transmits messages to several functional modules. For data loading, Image/Weights/Bias Reader operates independently but synchronizes at the end of each convolution instruction. These modules load data from on-chip buffers and transmit it to the AIE over various AXI streams. The ImageReader pads the data in accordance with the instruction information prior to transmitting it to AIE. The Bias Reader amalgamates the necessary bias data with runtime parameters essential for the ACC/NL, delivering critical runtime information, including the amount of accumulation times. Weights Reader only sends read requests to the Weight Buffer, which can handle data requests from multiple Convolution Engines and write directly into the AIE. The shared weight approach disseminates weights to all convolution processing elements to attain batch-level parallelism. The Image Writer rearrange the compute results from the AIE and transfer them back to the FM Buffer, ensuring they are stored in a coherent format for future operations.

\vspace{-0.05in}
\subsection{Low-Channel Convolution Unit}
\label{sec:low-channel-pe}
While the computational load of each model layer determines the overall inference latency, we observed a more significant decline in utilization during the computation of the inputs layer. Conv PE employs AIE cascade and feature map sharing to enhance the parallelism of the IC dimension, yielding low efficiency in calculating small channels, a frequent occurrence in the input layers of CV models. For instance, in ResNet50\cite{resnet}, stage 0 requires processing a convolution with a $7\times7$ kernel, IC=3, and OC=64. However, graph-level Conv PE has a channel parallelism of 64 (IC) $\times$ 128 (OC) (Section \ref{sec:convpe-graphlevel}). The efficiency is only 13.1\% when computing ResNet50's 1st later.

Combining batches can help traditional CNN accelerators achieve higher hardware utilization. However, AIE uses MAC chains to accumulate the IC dimension at runtime, which requires intricate data shuffling and substantially elevates control overhead. To address this challenge, DPUV4E chooses to deploy a configurable low-channel convolution unit on the PL side. The low-channel convolution unit's parallel parameters are specially designed for 1st layer dimension expansion, and the rest layers are proceed using AIE. Low-channel convolution unit and AIE can execute simultaneously, thereby mitigating the overhead of low-channel preprocessing. 
Compared to previous work \cite{xvdpu}, the substantial savings in DSP resources during the design phase enable us to implement this module.
In our experiments presented in Table \ref{tab:mlperf_result}, we employed a Low-Channel Convolution Unit with 4 (H) * 21 (IC) * 32 (OC) computational parallelism, requiring only $4\times21\times32/4 = 672$ DSP58s after employing INT8 DSP-packing\cite{dsp-packing-int8}. This led to a further performance boost of 793 FPS in addition to the $1.3\times$ improvement already attained with the normal 8PE architecture, while simultaneously decreasing latency by $7.5\%$ for a single batch.

%% file: 7.evaulation.tex
\section{Evaulation}
\vspace{-0.05in}
\subsection{Experiment Setup}
\label{sec:eval-setup}
We implemented DPUV4E on the VCK5000, a development board for data centers equipped with the Xilinx Versal ACAP XCVC1902. This development board is installed on a server using an AMD EPYC 7452 32-core Processor. We programmed the AIE, PL, and PS components using the Xilinx Vitis 2022.1 toolchain and used Vivado 2022.1 to generate the bitstream.

For the purpose of performance and resource evaluation, we chose the XVDPU (DPUCVDX8G) \cite{xvdpu}, which is also integrated into the VC1902 chip, alongside the DPUCZDX8G \cite{DPUCZDX8G} B4096-3-Core architecture introduced by Xilinx on the ZCU102. The B4096CU3 design, with a batch size of 3, demonstrates the highest level of parallelism possible in conventional FPGA. We also conducted a comparison with the latest CNN accelerator on FPGA\cite{h2pipe}. Despite XVDPU offering a C32B8 (batch size=8) design, considerable resource challenges arose during implementation; hence, we opted for the C32B6 (batch size=6) design in our comparison. 
All models were quantized using the Vitis AI toolchain.

\vspace{-0.05in}
\subsection{Implementation}
DPUV4E offers multiple configurations to accommodate FPGAs with varying resource capacities. The configurable units include numbers of PE (batch size) and specialized functionalities (normal, dwc, low-channel unit). 

Table \ref{tab:dpu-resource} illustrates some potential designs and their corresponding resource utilization. Compared to pure PL logic implemented on the ZCU102, the DPUV4E utilizing AIE conserves a substantial quantity of DSP slices and FFs, while delivering a minimum enhancement of $13.4\times$ in TOPS. 
Compared to the XVDPU, which similarly employs the VC1902 component, our design efficiently leverages idle AIE cores, facilitating the scheduling of up to 384 AIE cores (compared to 256 in the XVDPU) within constrained PL resources without compromising timing convergence (the XVDPU can only assure 300 MHz with 256 AIE cores).   
\input{tables/dpu_resource}

\input{tables/perf_all_v2}

\input{tables/mlperf_result_v2}

\vspace{-0.05in}
\subsection{End-to-end Performance}
Table \ref{tab:e2e-regular} presents the inference performance on various CNN models.
Through the efficient utilization of the AIE memory interface and data reuse, our 6PE design surpasses the performance of XVDPU under the same batch size while occupying fewer on-chip resources.
Specifically, our 6PE+DWC design attains a performance improvement of $1.14\times$ to $2.18\times$ compared to XVDPU. 

To verify the effectiveness of the DWC PE, we selected networks with DWC operations, such as the EfficientNet\cite{efficientnet}, MobileNet\cite{mobilenet,mobilenetv2}, and YOLOV5N\cite{yolov5}, to conduct end-to-end inference experiments. 
As illustrated in Table \ref{tab:e2e-regular}, compared to the conventional convolutional model (average improvement of 1.26$\times$), our design demonstrates a further performance enhancement with an average improvement of 1.78$\times$. This indicates that our designed DWC PE not only increases resource utilization for DWC operations but also effectively maintains compatibility with traditional convolutional operations.

\subsection{Compare with other accelerator}

To further demonstrate the performance comparison between DPUV4E and other designs\cite{nvidiaa100}\cite{nvidiaagx}\cite{nvidiat4}, we utilize the MLPerf\cite{mlperf} ResNet50\_v1.5 model to evaluate the actual inference performance.\footnote{Our design participated in the submission v1.0 of MLPerf Inference: Datacanter Benchmark. Detailed data can be found on the official website [URL] (hidden due to double-blind requirements)}
All models utilizing NVIDIA GPUs are quantised with TensorRT\cite{nvidiatensorrt} and evaluated on servers equipped with AMD EPYC 7742 CPUs. Models utilising the Xilinx FPGA are quantised with Vitis AI v1.4.1, and tests are performed according to the settings outlined in Section \ref{sec:eval-setup}.

Table \ref{tab:mlperf_result} illustrates the experiments we conducted using the same pre-trained model weights. In this table, FPS denotes the number of images processed per second during offline inference ($batch > 1$), while latency refers to the processing delay at $batch = 1$.

While FPGAs cannot match the elevated clock rates and computing capacity of ASICs, our solution attains similar low latency and high throughput while preserving generality.
Our design offers a $8.6\times/1.4\times$ improvement in TOPS/W compared to conventional FPGA-based DPU designs \cite{DPUCZDX8G} and the most recent AIE-based DPU designs \cite{xvdpu}. Compared with NVIDIA's T4 GPU based on the new generation NVIDIA TensorCore, it offers enhanced energy efficiency (TOPS/W) and superior throughput.
Compared to the NVIDIA Jetson AGX Xavier, our 8PE design can achieve high utilization under small batch conditions, offering up to $4.06\times$ the throughput improvement, while maintaining latency at only $81.5\%$ of the reference design.
In comparison to the data center accelerator card NVIDIA T4, our design increases throughput by $1.15\times$. The latency increases because the NVIDIA T4 offers $320.0$ GB/s GDDR6 bandwidth for real-time communication, while the VCK5000 development board we used only supports $102.4$ GB/s DDR4.
The single batch scenario cannot effectively leverage shared weights to mitigate the bandwidth limitation, resulting in higher latency.

Compared to the pure FPGA-based DPU B4096CU3, our design achieves a $37.02\times$ performance improvement. When compared to the current state-of-the-art FPGA+HBM CNN accelerator solutions\cite{h2pipe}, our design achieves a throughput enhancement of 6.23x. This is attributable to the powerful computational capabilities provided by the AI Engine.
In comparison to the AIE-based XVDPU 8PE at identical clock cycles and batch sizes, DPUV4E exhibits a $1.13\times$ performance improvement.  This could be attributed to the graph-level parallelism we offer, which ensures superior usage when the input/output channels are extensive. Our MAC chain cascade accumulation design also decreases element-wise accumulation time.
Our compact design and efficient module partitioning enable a maximum frequency that is $1.25\times$ that of XVDPU. In this scenario, the throughput is $1.35\times$ that of XVDPU.

To verify the effectiveness of our proposed low-channel convolution unit, we evaluated its performance in conjunction with the 8PE design(In the table it is denoted as 8PE+LowPE). The results indicate that the integration of the low-channel convolution unit into the design increases throughput by an additional $1.14\times$ and reduces latency by $7.54\%$.This innovation enables DPUV4E to enhance performance in latency-sensitive application contexts.

%% file: tables/dpu_resource.tex
\begin{table}[t]
\centering
\begin{threeparttable}[b]
  \setlength{\tabcolsep}{6pt} %
  \renewcommand{\arraystretch}{1} %
\caption{Resource utilization comparison}
\vspace{-0.05in}
\label{tab:dpu-resource}
\begin{tabular}{lcccccc}
\hline
\textbf{} &
  \textbf{\begin{tabular}[c]{@{}c@{}}DPUCZ\\ DX8G\cite{DPUCZDX8G}\end{tabular}} &
  \multicolumn{2}{c}{\textbf{\begin{tabular}[c]{@{}c@{}}XVDPU\\ \cite{xvdpu}\end{tabular}}} &
  \multicolumn{3}{c}{\begin{tabular}[c]{@{}c@{}}\textbf{DPUV4E}\\ \textbf{Ours}\end{tabular}} \\ \hline
\textbf{Device}    & ZCU102                                              & \multicolumn{2}{c}{VCK190} & \multicolumn{3}{c}{VCK5000}                                      \\ \hline
\textbf{Config}    & \begin{tabular}[c]{@{}c@{}}B4096\\ CU3\end{tabular} & C32B3        & C32B6       & \begin{tabular}[c]{@{}c@{}}6PE+\\ DWC\end{tabular} & 6PE  & 8PE  \\ \hline
\textbf{Freq(MHz)} & 281                                                 & 333          & 300         & 350                                                & 350  & 350  \\ \hline
\textbf{LUT(K)}       & 160                                                & 205         & 403        & 631                                               & 223 & 674 \\ \hline
\textbf{FF(K)}        & 315                                                & 266         & 507        & 648                                               & 287 & 696 \\ \hline
\textbf{BRAM}      & 771                                                 & 0            & 678         & 780                                                & 684  & 912  \\ \hline
\textbf{URAM}      & 0                                                   & 332          & 343         & 402                                                & 390  & 424  \\ \hline
\textbf{DSP}       & 1686                                                & 407          & 809         & 424                                                & 34   & 524  \\ \hline
\textbf{\begin{tabular}[c]{@{}l@{}}AIE\\ (MAC)\tnote{1}\end{tabular}} &
  \begin{tabular}[c]{@{}c@{}}0\\ (0)\end{tabular} &
  \begin{tabular}[c]{@{}c@{}}96\\ (96)\end{tabular} &
  \begin{tabular}[c]{@{}c@{}}192\\ (192)\end{tabular} &
  \begin{tabular}[c]{@{}c@{}}384\\ (192)\end{tabular} &
  \begin{tabular}[c]{@{}c@{}}384\\ (192)\end{tabular} &
  \begin{tabular}[c]{@{}c@{}}384\\ (256)\end{tabular} \\ \hline
\textbf{TOPS}      & 3.7                                                 & 32.7         & 61.4        & 72.5                                               & 68.4 & 91.2 \\ \hline
\end{tabular}
\begin{tablenotes}
\item[1] MAC refers to the number of AIEs utilized for performing Conv MAC operations.
\end{tablenotes}
\end{threeparttable}
\vspace{-0.2in}
\end{table}

%% file: tables/perf_all_v2.tex
\begin{table}
    \centering
      \setlength{\tabcolsep}{0.8pt} %
  \renewcommand{\arraystretch}{1.2} %
\label{tab:my-table}
  \caption {End-to-end FPS evaulation of different models on the DPUV4E}
\begin{tabular}{lccccccc}
\hline
\textbf{Model} &
  \textbf{\begin{tabular}[c]{@{}c@{}}Input \\ Shape\end{tabular}} &
  \textbf{GOPS} &
  \textbf{\begin{tabular}[c]{@{}c@{}}B4096\\ CU3\end{tabular}} &
  \textbf{\begin{tabular}[c]{@{}c@{}}XVDPU \\ C32B6\end{tabular}} &
  \textbf{\begin{tabular}[c]{@{}c@{}}Our \\ 6PE+DWC\end{tabular}} &
  \textbf{\begin{tabular}[c]{@{}c@{}}Our \\ 8PE\end{tabular}} &
  \textbf{Ratio} \\ \hline
\textbf{ResNet50\cite{resnet}}     & 3*224*224 & 8.19 & 190.3  & 2676.7 & 3417.8 & 4568.9 & 1.27 \\ \hline
\textbf{ResNet152\cite{resnet}}    & 3*224*224 & 21.8 & 84.7   & 1200.1 & 1586.1 & 2108.8 & 1.32 \\ \hline
\textbf{YOLOV3\cite{yolov3}}       & 3*416*416 & 65.9 & 37.5   & 286.8  & 382.9  & 472.2  & 1.33 \\ \hline
\textbf{SqueezeNet\cite{squeezenet}}   & 3*224*224 & 0.7  & 1500.8 & 5827.0 & 6658.9 & 7664.4 & 1.14 \\ \hline
\textbf{EfficientNet\cite{efficientnet}} & 3*224*224 & 4.7  & 319.0  & 2167.1 & 3976.5 & 3675.7 & 1.83 \\ \hline
\textbf{YOLOV5N\cite{yolov5}}      & 3*640*640 & 4.6  & 201.4  & 397.6  & 868.3  & 1379.8 & 2.18 \\ \hline
\textbf{MobileNetV1\cite{mobilenet}}  & 3*224*224 & 1.02 & 993.5  & 4913.3 & 8787.8 & 9123.1 & 1.78 \\ \hline
\textbf{MobileNetV2\cite{mobilenetv2}}  & 3*224*224 & 0.60 & 764.41 & 4930.3 & 6565.3 & 8315.8 & 1.33 \\ \hline
\end{tabular}
\label{tab:e2e-dwc}
\label{tab:e2e-all}
\label{tab:e2e-regular}
\vspace{-0.2in}
\end{table}

%% file: tables/mlperf_result_v2.tex
\begin{table*}[]
\caption{Performance comparison on MLPerf ResNet50\_1.5 }
\vspace{-0.05in}
\label{tab:mlperf_result}
  \setlength{\tabcolsep}{6pt} %
  \renewcommand{\arraystretch}{1.2} %
\begin{tabular}{lcccccccc}
\hline
\textbf{Type}    & GPU    & GPU    & H2PIPE\cite{h2pipe} & B4096CU3\cite{DPUCZDX8G} & XV-C32B8\cite{xvdpu} & Our 8PE & Our 8PE & Our 8PE+LowPE \\ \hline
\textbf{Processor} &
  \begin{tabular}[c]{@{}c@{}}NVIDIA Jetson\\ AGX Xavier\end{tabular} &
  \begin{tabular}[c]{@{}c@{}}NVIDIA T4\\ 16GB\end{tabular} &
  \begin{tabular}[c]{@{}c@{}}Intel\\ Stratix 10\end{tabular} &
  \begin{tabular}[c]{@{}c@{}}Xilinx ZU+ \\ MPSoC ZU9\end{tabular} &
  \begin{tabular}[c]{@{}c@{}}Xilinx ACAP \\ VC1902\end{tabular} &
  \begin{tabular}[c]{@{}c@{}}Xilinx ACAP\\ VC1902\end{tabular} &
  \begin{tabular}[c]{@{}c@{}}Xilinx ACAP\\ VC1902\end{tabular} &
  \begin{tabular}[c]{@{}c@{}}Xilinx ACAP\\ VC1902\end{tabular} \\ \hline
\textbf{TOPS}    & 32     & 130    & 7.73                & 3.7                      & 87.4                 & 131.0   & 163.8   & 165.7         \\ \hline
\textbf{Freq} &
  2.26GHz &
  585MHz &
  300MHz &
  300MHz &
  \begin{tabular}[c]{@{}c@{}}PL@300MHz\\ AIE@1.3GHz\end{tabular} &
  \begin{tabular}[c]{@{}c@{}}PL@300MHz\\ AIE@1.3GHz\end{tabular} &
  \begin{tabular}[c]{@{}c@{}}PL@375MHz\\ AIE@1.6GHz\end{tabular} &
  \begin{tabular}[c]{@{}c@{}}PL@375MHz\\ AIE@1.6GHz\end{tabular} \\ \hline
\textbf{Power} &
  \begin{tabular}[c]{@{}c@{}}30W\\ (TDP)\end{tabular} &
  \begin{tabular}[c]{@{}c@{}}70W\\ (TDP)\end{tabular} &
  - &
  \begin{tabular}[c]{@{}c@{}}25W(TDP)\\ 16.3W(Avg)\end{tabular} &
  \begin{tabular}[c]{@{}c@{}}125W(TDP)\\ 62.8W(Avg)\end{tabular} &
  \begin{tabular}[c]{@{}c@{}}125W(TDP)\\ 77.2W(Avg)\end{tabular} &
  \begin{tabular}[c]{@{}c@{}}125W(TDP)\\ 86.4W(Avg)\end{tabular} &
  \begin{tabular}[c]{@{}c@{}}125W(TDP)\\ 88.3W(Avg)\end{tabular} \\ \hline
\textbf{TOPS/W}  & 1.06   & 1.85   & -                   & 0.22                     & 1.39                 & 1.69    & 1.89    & 1.87          \\ \hline
\textbf{Latency} & 1.79ms & 0.83ms & 9.48ms              & 17.75ms                  & 1.97ms               & 1.75ms  & 1.46ms  & 1.35ms        \\ \hline
\textbf{\begin{tabular}[c]{@{}l@{}}FPS\\ (batch)\end{tabular}} &
  \begin{tabular}[c]{@{}c@{}}1346\\ (32)\end{tabular} &
  \begin{tabular}[c]{@{}c@{}}5423\\ (32)\end{tabular} &
  \begin{tabular}[c]{@{}c@{}}1004\\ (1)\end{tabular} &
  \begin{tabular}[c]{@{}c@{}}169\\ (3)\end{tabular} &
  \begin{tabular}[c]{@{}c@{}}4050\\ (8)\end{tabular} &
  \begin{tabular}[c]{@{}c@{}}4568\\ (8)\end{tabular} &
  \begin{tabular}[c]{@{}c@{}}5764\\ (8)\end{tabular} &
  \begin{tabular}[c]{@{}c@{}}6257\\ (8)\end{tabular} \\ \hline
\end{tabular}
\vspace{-0.1in}
\end{table*}

%% file: 8.conclusion.tex
\vspace{-0.1in}
\section{Conclusion}
In this paper, we propose DPUV4E, a high-performance inference accelerator based on Versal ACAP architecture. The DPUV4E attains a $95.8\%$ reduction in DSP Slices and a $44.7\%$ decrease in LUT resources by incorporating activation and non-convolution computations into the AIE MAC chain compared to previous works. This allows for the simple expansion of parallel batch sizes to $8$ and enables users to incorporate other custom logic.
Our approach optimally utilizes cascade connections, shared memory, and broadcast methods among AIEs for data reutilization. We develop two varieties of computation engines tailored for distinct computational patterns: Conv PE and DWC PE. Our technique yields an end-to-end performance enhancement of up to $2.1\times$ for models incorporating DWC and a $1.7\times$ improvement for alternative models, in contrast to prior works utilizing AIEs.
As an optional optimization strategy, we also offer a customizable Low-Channel Convolution Unit for data preprocessing to enhance the utilization of the DPUV4E’s compute array. Experiments on MLPerf ResNet50 indicate that our 8PE design maintains sufficient resources to implement this units, achieving a throughput improvement of 793 FPS.